\documentclass[epj,twocolumn]{webofc}
\usepackage[varg]{txfonts}   
\usepackage{color}
\woctitle{Powders \& Grains 2017}
\begin{document}
\title{Shear-induced organization of forces in dense suspensions: signatures of discontinuous shear thickening}
%
%

\author{\firstname{Sumantra} \lastname{Sarkar}\inst{1}\fnsep\thanks{\email{sumantra@mit.edu}} \and
        \firstname{Elan} \lastname{Shatoff}\inst{2}\fnsep \and
        \firstname{Kabir} \lastname{Ramola}\inst{2}\fnsep\and
            \firstname{Romain} \lastname{Mari}\inst{3}\fnsep\and
            	\firstname{Jeffrey } \lastname{Morris}\inst{4,5}\fnsep\and
            		\firstname{Bulbul} \lastname{Chakraborty}\inst{2}\fnsep\thanks{\email{bulbul@brandeis.edu}}
}
\institute{Physics of Living Systems, MIT, 400 Technology Square, Cambridge, MA 02139, USA
       \and
Martin Fisher School of Physics, Brandeis University, Waltham, MA 02454, USA           
\and
Department of Applied Mathematics and Theoretical Physics,
Centre for Mathematical Sciences, University of Cambridge, Cambridge CB3 0WA, UK
\and 
Benjamin Levich Institute, City College of New York, New York, NY 10031, USA
\and
Department of Chemical Engineering, City College of New York, New York, NY 10031, USA
}

\abstract{Dense suspensions can exhibit  an abrupt change in their viscosity in response to increasing shear rate.   The origin of 
this discontinuous shear thickening (DST) has been ascribed to the transformation of lubricated contacts to frictional, particle-on-particle contacts.  Recent research on the flowing and jamming behavior of dense suspensions has explored  the intersection of ideas from granular physics and Stokesian fluid dynamics to better understand this transition from lubricated to frictional rheology.  DST is reminiscent of classical phase transitions, and a key question is  how interactions between the microscopic constituents give rise to a 
macroscopic transition. In this paper, we extend a formalism that has proven to be  successful in understanding shear jamming of dry grains  to dense suspensions.   Quantitative analysis of the collective evolution of the contact-force network accompanying the DST transition demonstrates clear changes in the distribution of microscopic variables, and  leads to the  identification of   an ``order parameter''  characterizing DST.}
\maketitle
%

%

\section{Introduction}


A remarkable property of dense stabilized suspensions of particles in the tens of nanometers to tens of micrometer size range is that they can abruptly transform from a low to a high viscosity phase (or even a solid-like phase) with increasing applied stress~\cite{Brown2013,mari2014shear,Wyart_Cates,Waitukaitis:2012xy,Han:2016rt}.  Under steady shear,  these suspensions undergo a discontinuous shear thickening (DST)  transition~\cite{Brown2013,mari2014shear,Wyart_Cates}.    A confluence of ideas from the granular and  fluid dynamics communities has led to new understanding and a new set of questions regarding the flowing and jamming behavior of dense suspensions~\cite{Wyart_Cates,Cates_2014,GuazzelliMorris_2011}.   In this paper, we analyze the DST transition using a force-space representation~\cite{Sarkar:2016qy}  that was originally developed for granular systems.   We extend the formalism to suspensions and identify distinct, quantitative signatures of the DST transition in this representation.

There is developing consensus that particle contact and friction play 
a crucial role in dense suspension rheology~\cite{Brown_2012,Lootens_2005}.  Combining fluid mechanical interactions with contact friction  
between particles has been shown to capture critical features of both DST \cite{Seto_2013a,Mari_2014,Heussinger_2013, Fernandez_2013} and  shear induced jamming (SJ) \cite{Mari_2015}.  
Furthermore, clear connections have been made between a dynamic shear jamming front and impact-driven solidification~\cite{Han:2016rt}.   
The appearance of non-monotonic flow curves in numerical models of DST~\cite{Seto_2013a,Mari_2014} is reminiscent of classical phase transitions.   As in classical phase transitions, the key question is  how interactions between the microscopic constituents give rise to a 
macroscopic transition.    

The essential idea underlying the jamming-based rheology model~\cite{Wyart_Cates,Cates_2014} for DST is that it requires a shear-stress driven transition from the lubricated to the frictional branch of the viscosity.    Reduction of  the jamming packing fraction  through the creation of frictional contacts~\cite{GuazzelliMorris_2011} yields a higher viscosity at any $\phi$.   If shearing changes the relative fractions of frictional and lubricated contacts, the suspension can transition to the frictional branch.  {We are interested in understanding the nature of the collective reorganization of particles that gives rise to this change in rheology.}
Simulations show that DST is accompanied by  significant changes in the  network of frictional contacts and contact forces with minimal changes in  structure factors and pair correlation functions~\cite{mari2014shear}.   We present a quantitative analysis of  changes in the force network by studying the organization in  ``force space'', which is dual to the positional network of grains in a sense to be defined below~\cite{Sarkar:2016qy,Sarkar:2013zr}.  {The density of points in this space naturally decreases as the stress increases since the number of contacts is roughly constant, but more importantly we show that  the rheological  behavior characterizing  DST is accompanied by changes in the form of the density distribution of points in this space.   The change in distribution is a consequence of  the constraints of local mechanical equilibrium applied to grains as contacts change from  lubricated to frictional.} 
%
\section{Model}
The combination of frictional and viscous interactions between surfaces of suspended particles has recently been addressed
through simulations~\cite{Seto_2013a,Mari_2014, Heussinger_2013, Fernandez_2013} that use a hybridization~\cite{Seto_2013a,Mari_2014} of  Stokesian Dynamics (SD) with discrete-element modeling (DEM)~\cite{Cundall_1979}. The SD method is simplified by only considering the near-field, or lubrication, hydrodynamic interactions.
To allow contact,  the lubrication resistance singularity  is cut off at a distance $h_c$ between particle surfaces.
The DEM approach uses a Coulomb friction law: for $F_{\tan} \le \mu F_{norm}$, where $\mu$ is the interparticle friction coefficient,
there is no slipping of the contact, while there is slip when the friction is fully mobilized for $F_{\tan} = \mu F_{norm}$.
To model the transition from lubricated to frictional contacts,  a repulsive force, representative of electrostatic repulsion or steric hindrance due to a grafted layer, is used.  For low stress or low shear rate, the repulsive force maintains surfaces separated at $h>h_c$ and thus interactions are lubricated, as if $\mu = 0$, while at large enough stress to push particles to $h<h_c$,  frictional contacts are activated.  
2D simulations of this model capture the progressively steeper shear thickening with increasing $\phi$ (Fig. \ref{fig::point_pattern}a) seen in experiments.  


\section{Force Balance in 2D Suspensions}
Owing to the size ($\le 10 \mu m$) of the particles and the shear rates at which DST is observed
(usually around $1 s^{-1}$), it is a very low Stokes number phenomenon, and particle inertia can be neglected.
As a consequence, the constraints of mechanical equilibrium are strictly obeyed at both the local and global levels.
We analyze  data obtained from stress-controlled  simulations of  a bidisperse suspension of particles using the  hybrid SD-DEM model~\cite{Seto_2013a,Mari_2014} in 2D.
Each disk in the suspension experiences a hydrodynamic force, $\vec F_H$,
a short range repulsive force, $ \vec F_R $, and frictional contact force, $\vec F_C $.
The hydrodynamic component includes a non pairwise force (Stokes drag) and pairwise forces (lubrication),
whereas $\vec F_R$ and $\vec F_C$ are pairwise interactions.
The force balance equation for each disk is $\vec F_H + \vec F_R +  \vec F_C = 0$.
The repulsive force is modeled as a short range electrostatic repulsion, 
\begin{equation}\label{repulsion}
 \vec F_R^{(i,j)} \propto \begin{cases}
- \frac{a_ia_j}{a_i + a_j}  e^{-\kappa h^{(i,j)}} & h^{(i,j)} \geq 0 \\
- \frac{a_ia_j}{a_i + a_j}   & h^{(i,j)} \leq 0 ~,\\
\end{cases}
\end{equation}
with Debye length, $ 1/\kappa $, much smaller than the radius of  either particle, $a= ( a_i$ {\rm or} $ a_j )$.  Adjacent grains that overcome this repulsive force form frictional contacts and contribute to the $ \vec F_C $. Thus, $|F_R|$  sets a stress scale, $\sigma_0 \approx \frac{|F_R|}{a^2}$, whose interplay with the hydrodynamic stress scale determines the stress (shear rate) dependent rheology. 

\subsection{Force tilings in granular materials}
\begin{figure}
	\centering
	\includegraphics[width=0.5\textwidth] {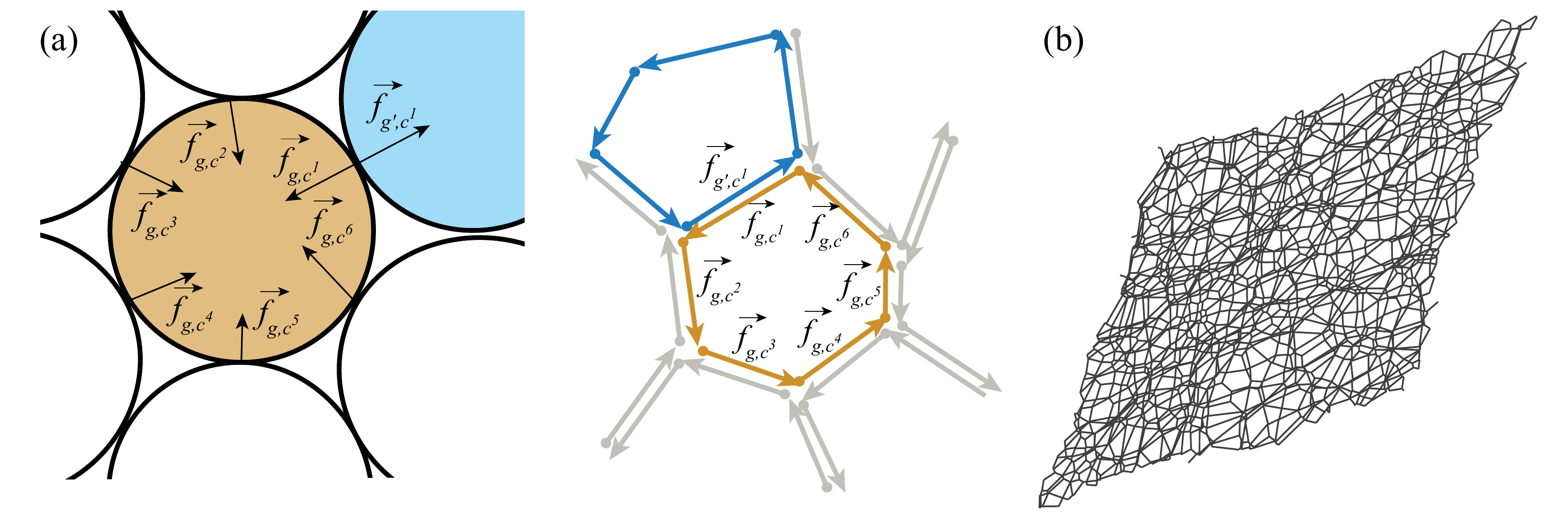}
	\caption{
				(a) Construction of force tiles: The vectors $\vec{f}_{g,c}$ form a closed polygon (light brown) (Eq. \ref{force_sum_zero}),  and there is strict edge sharing between grains $g$ and $g^{\prime}$ (Eq. \ref{second_law}).  
		(b) A force tiling constructed from a configuration at $ \sigma=\sigma_{xy}/\sigma_0 = 0.1 $. The shape and size of the bounding parallelogram is determined by the components of the imposed stress tensor: $\sigma$ controls the opening angle.  The number of tiles is equal to the number of particles in the force-bearing network}
	\label{fig::height} 
\end{figure}

In packings of dry grains,  labeling the  contacts of grain $g$ by  $(g, c)$ (Fig. \ref{fig::height}),  the force balance condition is:
 \begin{equation}
\vec F_C \equiv \sum_{c} \vec{f}_{g,c} = 0,
\label{force_sum_zero}
\end{equation}
where the sum is taken over all the contacts $\{c\}$ for a given grain $g$.
In addition, Newton's {\it third law} states:
\begin{equation}
\vec{f}_{g,c} = -\vec{f}_{g',c}.
\label{second_law}
\end{equation}
These constraints  can  be used  to construct a  representation  known as the ``force tile" representation or Maxwell-Cremona tiles~\cite{Brian_Jstat}. Eq. (\ref{force_sum_zero}) can be represented
as a {\it closed polygon}, 
if the sum is taken {\it cyclically} over the contacts for each grain.
Eq. \ref{second_law}  dictates that the tiles of touching grains have to share an edge.   We can, therefore,  construct a {\it force tiling}, as illustrated in Fig. \ref{fig::height}.
The vertices of the force tiling form a point pattern.   
In previous studies of the shear-jamming process in dry granular systems, it has been shown that the onset of shear-jamming is marked by distinct changes in this point pattern~\cite{Sarkar:2016qy,Sarkar:2013zr}.

\subsection{Generalization to Suspensions} 
The Stokes drag is a non-pairwise force that prevents  individual force tiles from closing: $\sum_{c} \vec{f}_{g,c} = \vec{f}_{Stokes}$. In addition,  suspended grains experience pairwise, non-contact forces.  The latter can be included by extending the definition of  the  contact network, $\lbrace g, c \rbrace$, and $\lbrace \vec{ f_{g,c}} \rbrace$ to  include $\vec F_R$, $\vec F_c$ and the pairwise lubrication force.   Given these, and $\vec{f}_{Stokes}$,   there is a unique way of obtaining effective contact forces such that all force tiles close~\cite{kabir_kadanoff}.    We use an iterative algorithm~\cite{Sarkar:2016qy} in which the contact forces are modified to ensure that  Eqs. \ref{force_sum_zero} and \ref{second_law} are satisfied at every iteration. The final solution is unique up to global translation.

\section{Results}
\begin{figure}[htbp]
	\centering
	\includegraphics[width=0.5\textwidth] {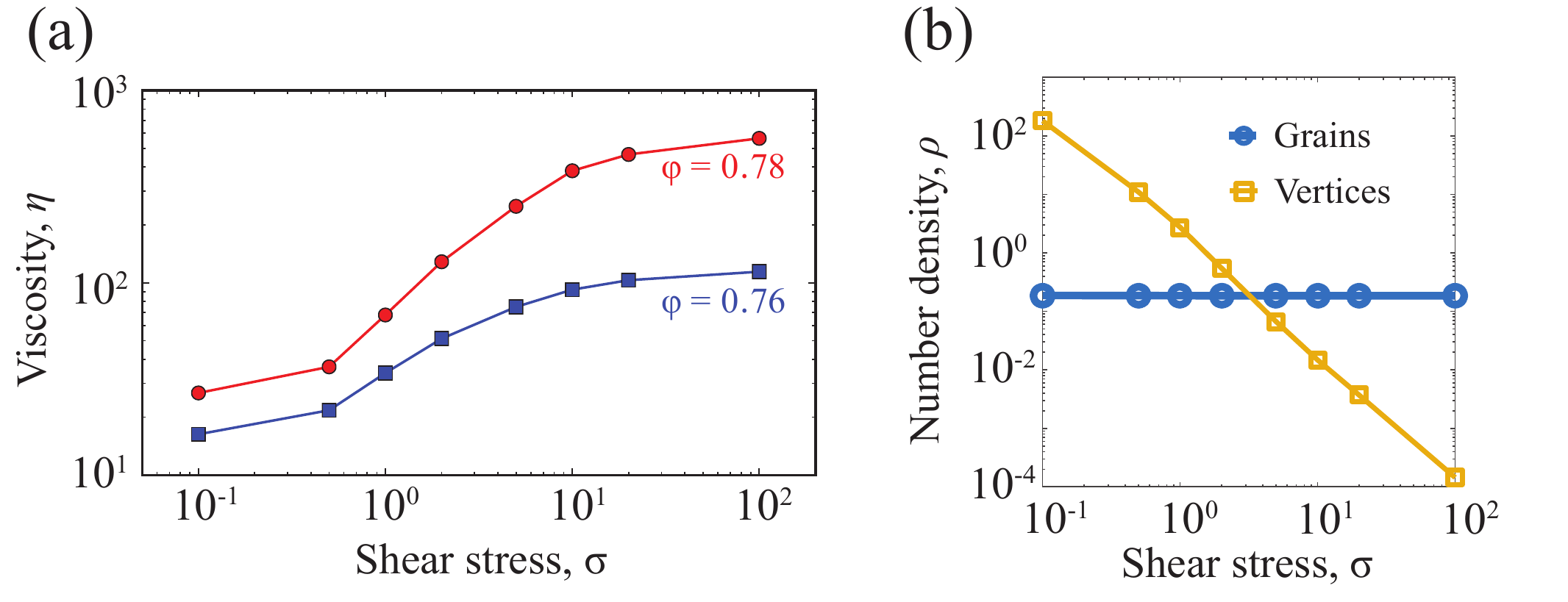}
	\caption{(a) Viscosity as a function of shear stress in  stress-controlled simulations of a non-Brownian suspension. (b) Density of grains and vertices as a function of shear stress. }
		
	\label{fig::point_pattern} 
		  \end{figure}

We construct force tilings for each member of the ensemble of steady states over a range of  shear stresses, $\sigma \equiv \sigma_{xy}/\sigma_0$ and at packing fractions, $\phi=0.76$ and $0.78$.   
The point patterns of force-tiling vertices from   three representative configuration at $\sigma$ below and above the DST transition for $\phi=0.78$ are shown in Fig. ~\ref{fig::point_pattern}. The lengths of the boundaries increase with increasing $\sigma$, and they have been scaled by $\sigma$ in order to aid in the visualization of the patterns.  It should be pointed out that the shape of the bounding boxes fluctuate even at a given $\sigma$, which fixes only the opening angle.  Fig. ~\ref{fig::point_pattern} (b) illustrates the effect of increasing  $\sigma$ on the density of vertices.  This density is observed to follow $\rho_v \propto  1/\sigma^2$ for up to six decades, and occurs as $\phi$ is held constant with no  change in the real-space density. 
Given the dramatic change in $\rho_v$ across the DST, one could expect significant changes in the {\it patterns} of vertices.    We have employed a clustering analysis of these point patterns based on purely local densities.  In the next section we  demonstrate that there is a clear signature of the DST transition in the clustering properties of the vertices.   

\subsection{Point Patterns of Vertices}
\subsubsection{Density based clustering analysis}
{If the repulsive force between two contacts is overcome to create a frictional contact, the constraints of mechanical equilibrium will necessarily induce a change in the shape of the force tiles, and hence a collective reorganization of the point pattern of vertices.  We performed a density based clustering analysis of the vertices of force tiles using the DBSCAN algorithm~\cite{Ester96adensity-based} to detect these changes. In DBSCAN, 
two points belong to the same cluster if their distance is less than a probing length scale, $l$ (which  here will dimensionally be a stress scale). The connected set of all such points define a unique cluster, which  are separated from each other by distances greater than $l$.
%
%
%
In usual implementations of DBSCAN,  clusters are determined  using an optimum  probing radius.  

We use DBSCAN to analyze how the clustering pattern changes as the probing length scale is varied.  For a large enough length scale, all points will belong to one cluster and for small enough length scales, each point will be its own cluster.  
Our algorithm probes the density distribution of point patterns at different length scales by monitoring the  number of clusters,  $N_c(l)$,  as a function of the probe  length, $l$. For a point pattern with uniform density, $N_c(l)$ decreases continuously with $l$.  In a periodic lattice, where the distance distribution of nearest neighbors is a delta function,  $N_c(l)$ exhibits a jump discontinuity at  the lattice spacing. For a complex pattern, we expect $N_c(l)$ to show significant changes in its derivatives at length scales where the distance distribution has structure.}

\begin{figure}[htbp]
	\centering
	\includegraphics[width=0.45\textwidth] {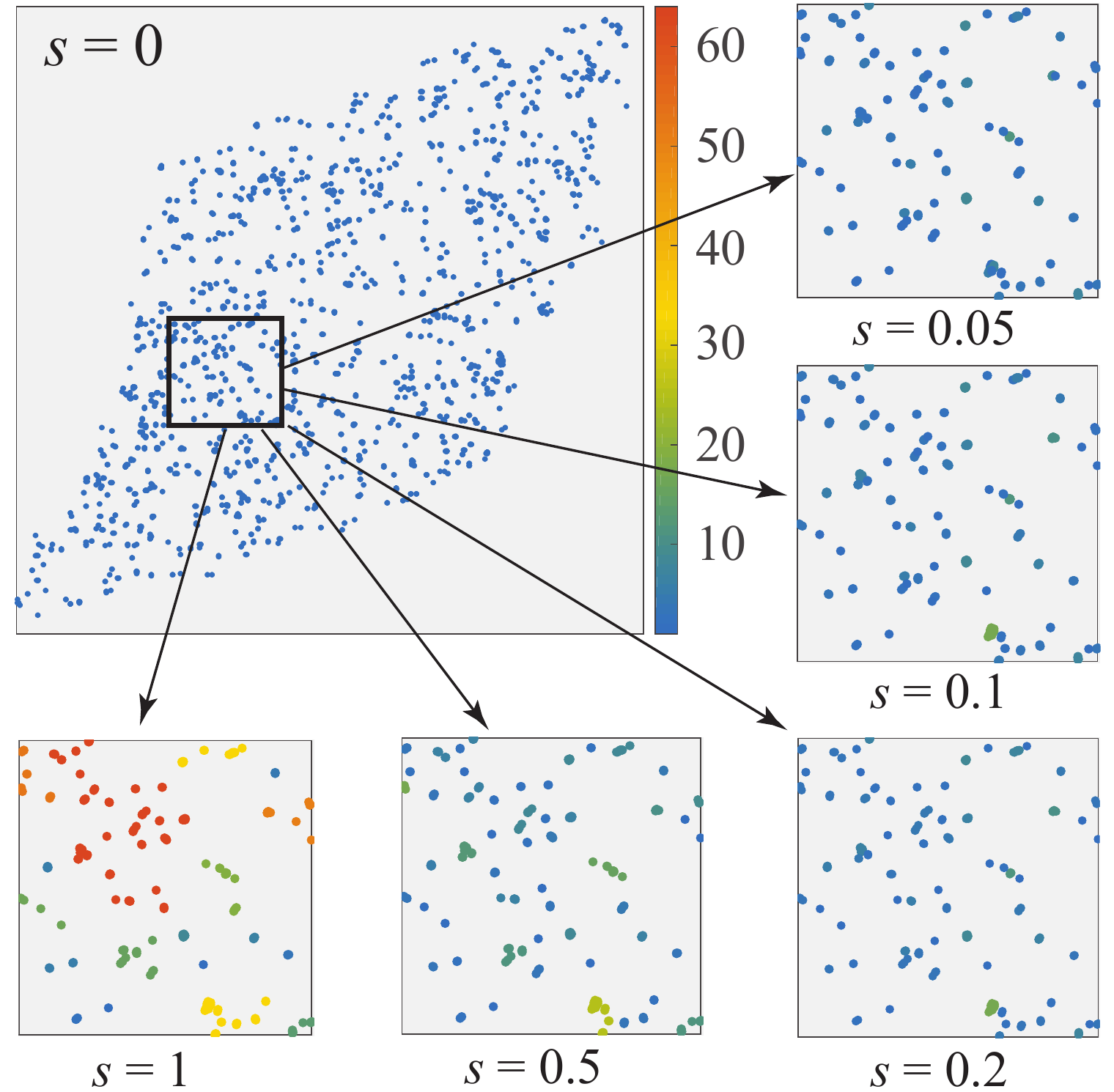}
	\caption{Illustration of DBSCAN analysis.  Left: Sample tiling at $\phi=0.78$, $\sigma=100$ at $s=0$:  all clusters are of size $1$ and $N_c(s=0) = N_v$, the number of vertices.  Right panel:  Zoomed in point pattern analyzed with increasing probe length $s$.  The clusters are colored by their size as indicated by the color bar.}
	\label{DBSCAN} 
\end{figure}In applying the clustering analysis to point patterns in force tilings, the probing  ``length scale''  is to be interpreted as an isotropic probing stress scale, $\sigma_{probe} $.   We measure the variation in the number of clusters $N_c(s)$ as the dimensionless probe stress, $ s = \sigma_{probe}/\sigma$,   is increased (Fig. \ref{DBSCAN}).  Identifying significant variations in $ N_c(s) $ yields characteristic stress scales in the patterns,  just as in real-space. 
 
\subsubsection{Clustering Analysis of DST} 
\begin{figure}[htbp]
	\centering
	\includegraphics[width=0.5\textwidth] {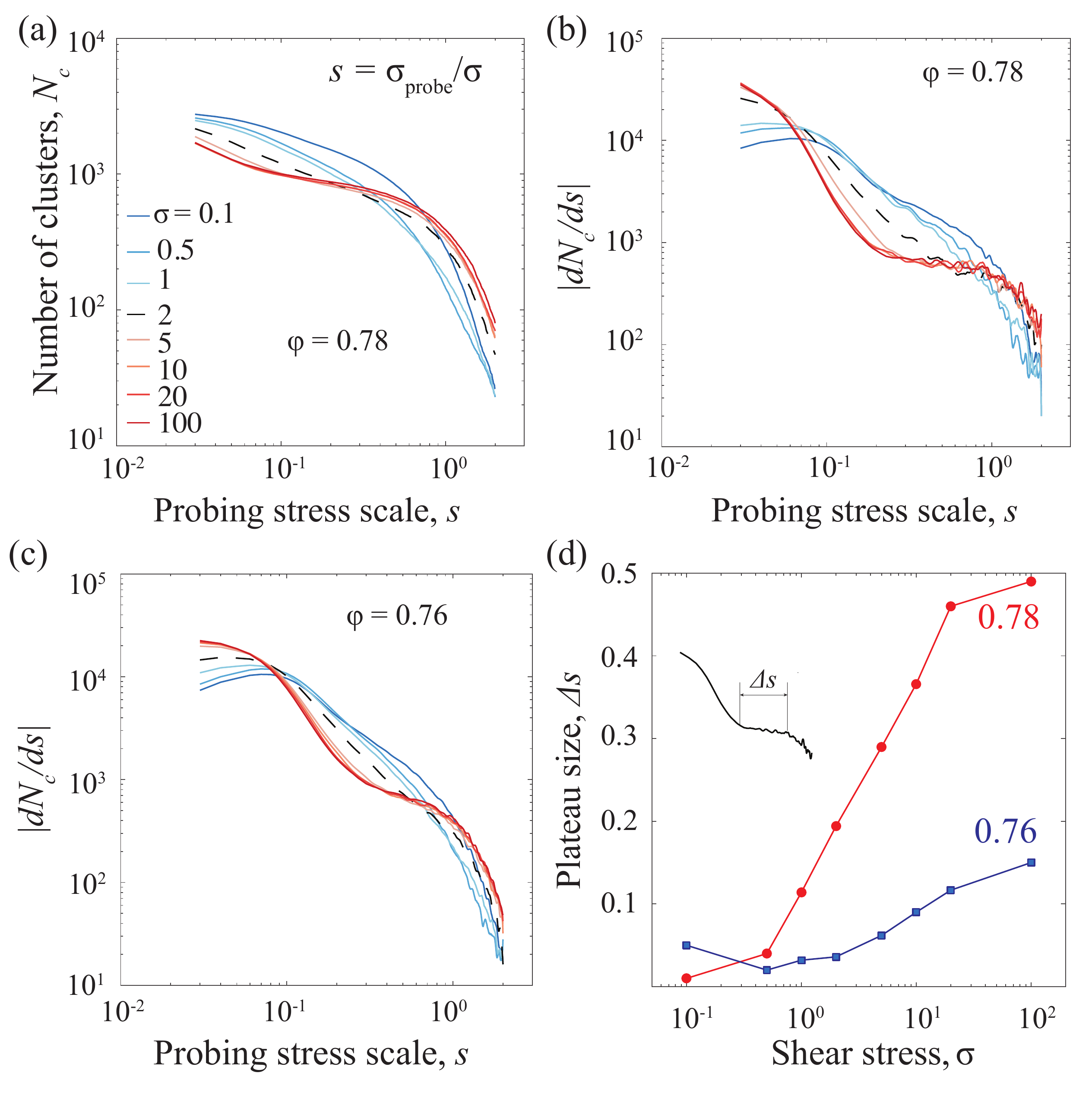}
	\caption{Top: Cluster analysis at $\phi=0.78$: Number of clusters, $ N_c $,  versus $ s = \sigma_{\mbox{probe}}/\sigma$, at different values of $\sigma$.   b) $ dN_c/ds $ vs $ s $.   Each curve is obtained by averaging over fifty steady state configurations.  The black dashed lines highlight the behavior change at $\sigma \approx 2$. Bottom:   (c) $ dN_c /ds$ vs $ s $ for $ \sigma = 0.1 \mbox{ and }100 $ at $ \phi = 0.76 $ compared to  $ \phi = 0.78 $.  (d) Plateau size $ \Delta s $ (See inset for the definition) as a function of the control shear stress $ \sigma$ for packing fractions 0.76 and 0.78 . }
	\label{fig::cluster_analysis} 
\end{figure}

We calculate $N_c(s)$ for each force tiling and then ensemble average over tilings sampled at a given $\phi$ and $\sigma$.
Fig. ~\ref{fig::cluster_analysis} (a) shows that at $\phi=0.78$, and $\sigma$ below the DST transition, $N_c(s)$ decays continuously with $s$, indicating the lack of any characteristic scale in the distribution of distances between the vertices.  We note that vertices that are close to each other are not necessarily connected by an edge in the tiling.  Therefore, the distance distribution is not equivalent to the contact force distribution.

Above the DST transition, we can clearly identify three different decay regimes in $N_c (s)$, with a plateau-like  structure clearly identifiable in the derivative, $dN_c(s)/ds$ shown in Fig. ~\ref{fig::cluster_analysis} (b). As seen from Fig. \ref{fig::cluster_analysis} (c), this structure in  $dN_c(s)/ds$ is much more pronounced for $\phi=0.78$ than it is for $\phi=0.76$.   The observed structure in $N_c(s)$ indicates that there are density inhomogeneities in the pattern of the vertices characterized by both $s^*$, marking the onset of the plateau, and $s^{**}$, the edge of the plateau.  The abrupt change in $dN_c(s)/ds$ to a much smaller value at $s^*$ implies that incremental changes in $s$ beyond $s^*$ incorporate the ``noise points'':  isolated vertices that lie in low-density regions.   Beyond $s^{**}$, $dN_c(s)/ds$ starts decreasing as $s$ increases indicating that  clusters separated by $s \ge s^{**}$ are beginning to merge.   Fig. \ref{DBSCAN} illustrates how the cluster pattern changes with $s$.  The difference $\Delta s= s^{**}-s^*$ provides a measure of the separation of high density regions in force tilings: the more pronounced the plateau the sharper is the distinction between high and low density regions. 

Fig. \ref{fig::cluster_analysis} (d) demonstrates that  $\Delta s$ increases sharply at a characteristic value of $\sigma$ that depends on $\phi$.  Moreover the saturation value of $\Delta s$ decreases with decreasing $\phi$, suggesting that $\Delta s$ can  be considered as an ``order parameter'' characterizing DST.
%
\section{Discussion}
{Using a clustering analysis, we have demonstrated that the rheological changes at DST are accompanied by a collective reorganization in the space of forces.
%
%
This collective response is necessitated by the constraints of mechanical equilibrium, applied at the local grain level, as  lubrication forces  change to frictional contact forces.
We identify a characteristic stress scale in the  pattern, $\Delta s$, from  the plateau in $dN_c(s)/ds$.   This plateau suggests that the vertices are clustered into clumps with an approximately uniform density of points with the  clumps separated by $\approx \Delta s$.  Equilibrium clumped phases are observed in systems with  ultrasoft interaction potentials~\cite{PhysRevE.94.042120,KLEIN1994738}.  Indeed, our preliminary DBSCAN analysis of the clumped phase shows a plateau structure remarkably similar to Fig. \ref{fig::cluster_analysis} (b).   We are currently exploring this connection further}.
%
\section*{Acknowledgements and References} 
The authors acknowledge useful conversations with Jetin Thomas and Abhinendra Singh.   SS, KR and BC acknowledge  the support of NSF-DMR 1409093 and the W. M. Keck Foundation.  JM and BC  acknowledge NSF-CBET 1605283.

%

%
%
%

\end{document}